\documentclass{jfm}

\usepackage{graphicx}
\usepackage{amsmath}
\usepackage{amssymb}
\usepackage{pifont}
\usepackage{txfonts}
\usepackage{psfrag}
\usepackage[round]{natbib}
\usepackage[T1]{fontenc}
\usepackage{lmodern}
\usepackage{tikz}
\usetikzlibrary{shapes.geometric,shapes.symbols}
\usepackage{scalerel}
\usepackage{upgreek}
\usepackage{booktabs}
\usepackage{verbatim}
\usepackage{framed}
\usepackage{fancyvrb}
\usepackage{varwidth}
\usepackage{bbding}
\usepackage{subfigure}
\usepackage{multirow}
\usepackage[normalem]{ulem}
\usepackage{multicol}
\usepackage{xcolor}
\usepackage{breqn}
\usepackage{soul}
\usepackage{dirtytalk}
\usepackage{upquote}
\usepackage{authblk}
\usepackage{array}
\newcolumntype{P}[1]{>{\centering\arraybackslash}p{#1}}

\title{Active and inactive components of the streamwise velocity in wall-bounded turbulence}

\shorttitle{Active and inactive components of the streamwise velocity in wall turbulence}

\shortauthor{R. Deshpande, J. P. Monty and I. Marusic}

\author[1]{Rahul Deshpande}
\author[1]{Jason P. Monty}
\author[1]{Ivan Marusic\thanks{Email address for correspondence: imarusic@unimelb.edu.au}}
\affil[1]{Department of Mechanical Engineering, University of Melbourne, Parkville, VIC 3010, Australia}

\pubyear{2020}
\date{?; revised ?; accepted ?. - To be entered by editorial office}
\begin{document}

\maketitle

\begin{abstract}
\citet{townsend1961} introduced the concept of active and inactive motions for wall-bounded turbulent flows, where the active motions are solely responsible for producing the Reynolds shear stress, the key momentum transport term in these flows. 
While the wall-normal component of velocity is associated exclusively with the active motions, the wall-parallel components of velocity are associated with both active and inactive motions. 
In this paper, we propose a method to segregate the active and inactive components of the 2-D energy spectrum of the streamwise velocity, thereby allowing us to test the self-similarity characteristics of the former which are central to theoretical models for wall-turbulence.
The approach is based on analyzing datasets comprising two-point streamwise velocity signals coupled with a spectral linear stochastic estimation (SLSE) based procedure. 
The data considered span a friction Reynolds number range $Re_{\tau}$ $\sim$ ${\mathcal{O}}$($10^3$) -- ${\mathcal{O}}$($10^4$).
The procedure linearly decomposes the full 2-D spectrum (${\Phi}$) into two components, ${\Phi}_{ia}$ and ${\Phi}_{a}$, comprising contributions predominantly from the inactive and active motions, respectively.
This is confirmed by ${\Phi}_{a}$ exhibiting wall-scaling, for both streamwise and spanwise wavelengths, corresponding well with the Reynolds shear stress cospectra reported in the literature.
Both ${\Phi}_{a}$ and ${\Phi}_{ia}$ are found to depict prominent self-similar characteristics in the inertially dominated region close to the wall, suggestive of contributions from Townsend's attached eddies.
Inactive contributions from the attached eddies reveal pure $k^{-1}$-scaling for the associated 1-D spectra (where $k$ is the streamwise/spanwise wavenumber), lending empirical support to	the attached eddy model of \citet{perry1982}.
\end{abstract}

\keywords{boundary layer structure, turbulent boundary layers, turbulent flows}

\section{Introduction and motivation}
\label{intro}

The attached eddy model \citep{perry1982,marusic2019}, based on Townsend's attached eddy hypothesis \citep{townsend1976}, is a conceptual model through which the kinematics in a wall-bounded flow can be statistically represented by a hierarchy of geometrically self-similar attached eddies that are inertially dominated (inviscid), and randomly distributed in the flow field.
Here, the term `attached' refers to a flow structure whose geometric extent, i.e. the size of its velocity field, scales with its distance from the wall ($z$) and mean friction velocity ($U_{\tau}$), respectively.
As per \citet{townsend1976}, the attached eddies have a population density inversely proportional to their height ($\mathcal{H}$), which ranges between ${\mathcal{O}}(z_{min})$ $\lesssim$ $\mathcal{H}$ $\lesssim$ ${\mathcal{O}}(\delta)$, where $z_{min}$ corresponds to the start of the inertial region, while $\delta$ is the boundary layer thickness.
At any $z$ $\gtrsim$ $z_{min}$, the cumulative contribution from the range of attached eddies results in the streamwise and spanwise turbulence intensities varying logarithmically as a function of $z$, while the wall-normal variance is a constant following:
\begin{equation}
\label{eq1}
\begin{aligned}
{{\overline{u^2}}^+} &= {B_1} - {A_1}\; \ln(\frac{z}{\delta}), \\
{{\overline{v^2}}^+} &= {B_2} - {A_2}\; \ln(\frac{z}{\delta}), \\
{{\overline{w^2}}^+} &= {B_3},\; \text{and} \; {{\overline{uw}}^+} = {B_4},
\end{aligned}
\end{equation}
where $A_1$, $A_2$, $B_1$, $B_2$, $B_3$ and $B_4$ are constants. 
Here, $u$, $v$ and $w$ are the velocity fluctuations along the streamwise ($x$), spanwise ($y$) and wall-normal ($z$) directions, respectively, while superscript `+' denotes normalization by $U_{\tau}$ and kinematic viscosity ($\nu$).
Recent literature \citep{jimenez2008,baidya2014,mklee2015,orlandi2015} has reported substantial support for expressions corresponding to the lateral velocity statistics in (\ref{eq1}), from experimental as well as simulation data, down to as low as $z^{+}$ $\sim$ 100. 
Support for a log law for ${\overline{u^2}}^{+}$ has been more convincing from high $Re_{\tau}$ experimental datasets \citep{hultmark2012,marusic2013} in comparison to low $Re_{\tau}$ simulations \citep{jimenez2008,mklee2015}, likely owing to the lack of scale separation resulting in the self-similar contributions becoming obscured by the non-self-similar contributions at the same scale \citep{jimenez2008,rosenberg2013,baars2020part2}.
Recently, \citet{baars2020part2} were able to segregate these two contributions, consequently revealing the near-wall logarithmic growth (of ${\overline{u^2}}^{+}$) due to self-similar contributions down to $z^{+}$ $\sim$ 80, with a slope of 0.98 ($=$ $A_{1}$; also known as the Townsend-Perry constant).

Given that the turbulence intensities in (\ref{eq1}) equate to the integrated spectral energy in the respective velocity fluctuations (that is, $\overline{u^2}$ $=$ $\int_{0}^{\infty} {{\phi}_{uu}} {d{k_x}}$, where ${\phi}_{uu}$ is the one-dimensional (1-D) streamwise velocity spectrum and $k_{x}$ is the streamwise wavenumber), the contribution from the hierarchy of attached eddies also manifests itself in the energy spectra of the two wall-parallel velocity components; in the form of the so-called $k^{-1}_{x}$-scaling \citep{perry1982}.
This scaling has been predicted previously via dimensional analysis and other theoretical arguments \citep{perry1977,perry1986,nikora1999,katul2012}, with \citet{perry1986} further arguing that the respective premultiplied spectra (${k^{+}_{x}}{{\phi}^{+}_{uu}}$, ${k^{+}_{x}}{{\phi}^{+}_{vv}}$) should plateau at a constant value equal to the rate of logarithmic decay ($A_{1}$ and $A_{2}$) for ${\overline{u^2}}^{+}$ and ${\overline{v^2}}^{+}$. 
These predictions, however, are rarely observed at finite $Re_{\tau}$, likely due to the flow containing a mixture of self-similar attached eddies and other non-self-similar flow structures.
The difficulty in separating the two contributions may explain the lack of convincing empirical evidence of the $k^{-1}_{x}$-scaling for ${\phi}_{uu}$, and its association with $A_1$, in the literature \citep{nickels2005,rosenberg2013,baars2020,baars2020part2}.

Noting that experiments show that $\overline{u^2}^+$ and $\overline{v^2}^+$ varies with Reynolds number in the inertial region while $\overline{uw}^+$ does not (as per equations \ref{eq1}), \citet{townsend1961} commented that ``\emph{it is difficult to reconcile these observations without supposing that the motion at any point consists of two components, an active component responsible for turbulent transfer and determined by the stress distribution and an inactive component which does not transfer momentum or interact with the universal component.}''
He further elaborated ``\emph{that the inactive motion is a meandering or swirling motion made up from attached eddies of large size which contribute to the Reynolds stress much further from the wall than the point of observation.}''
This definition of active and inactive motions, however, seems to have been interpreted differently by some in the literature. 
Therefore, we attempt to clarify here our (and Townsend's) interpretation and emphasise its consistency with the attached eddy hypothesis (AEH). 

\subsection{Active and inactive motions}

In the simplest attached eddy model, attached eddies are the only eddying motions present in the boundary layer, and they lead to `active' and `inactive' contributions. 
The key reason for this is the nature of the velocity signature from individual attached eddies in this inviscid model. 
The impermeability boundary condition at the wall enforces $w=0$ at the wall, but allows slip (and hence finite $u$ and $v$ at the wall). This is achieved by producing attached eddy velocity fields using a vortex structure with image vortex pairs in the plane of the wall. The result is a spatially localised $w$-velocity signature from the attached eddies - this is well illustrated in figure 1 of \citet{perry1986}. 
Consequently, at any wall-normal location $z$ in the inertial region, active motions are solely due to the velocity fields of the attached eddies of height, $\mathcal{H}$ $\sim$ $\mathcal{O}$($z$), and these contribute to $u(z)$, $v(z)$, $w(z)$ and hence ${\overline{uw}}$($z$). 
The inactive motions, however, are caused by the velocity fields from relatively large and taller attached eddies of height $\mathcal{O}$($z$) $\ll$ $\mathcal{H}$ $\lesssim$ $\mathcal{O}$($\delta$), and while these eddies contribute to $u(z)$ and $v(z)$, they make no significant contribution to $w(z)$. Hence, the inactive motions do not contribute to ${\overline{uw}}$($z$) (or $\overline{w^2}$($z$)).
Therefore, while both active and inactive motions contribute to ${\overline{u^2(z)}}$ (and ${\overline{v^2(z)}})$, there are only active contributions to ${\overline{uw}}$($z$) (or $\overline{w^2}$($z$)). 
The consequence of this is that active motions are the component of attached-eddy contributions that have pure wall-scaling ($z$ and $U_\tau$).
The remaining attached eddy contributions are the relatively large scale inactive motions which, together with the inverse probability distribution of scales as per AEH, lead to the logarithmic decay of ${\overline{u^2}}^{+}$ and ${\overline{v^2}}^{+}$ (equation \ref{eq1}) with $z$.

Given the above, the resulting attached eddy velocity fields can thus be decomposed following \citet{panton2007}:
\begin{equation}
\label{eq2}
\begin{aligned}
u &= {u}_{\rm active} + {u}_{\rm inactive}, \\
v &= {v}_{\rm active}+ {v}_{\rm inactive}, \\
w &= {w}_{\rm active},
\end{aligned}
\end{equation}
and as the active and inactive velocity fields are uncorrelated \citep{townsend1961,bradshaw1967}, the Reynolds stresses in equation (\ref{eq1}) can also be decomposed as:
\begin{equation}
\label{eq3}
\begin{aligned}
\overline{u^2} \; &= \; \overline{{u}^{2}}_{\rm active} + \overline{{u}^{2}}_{\rm inactive},\\
\overline{v^2} \; &= \; \overline{{v}^{2}}_{\rm active} + \overline{{v}^{2}}_{\rm inactive},\\
\overline{w^2} \; &= \; \overline{{w}^{2}}_{\rm active}, \\
\overline{uw} \; &= \; \overline{({u}_{\rm active})({w}_{\rm active})}.
\end{aligned}
\end{equation}
Here, the active and inactive motions can be deemed uncorrelated only if we ignore the non-linear interactions across these motions, such as modulation, which have been shown to exist previously \citep{morrison2007,mathis2009,marusic2010,chernyshenko2012,WuChristensenPantano2019}.
However, such interactions will not contribute significantly to second-order velocity statistics (equation \ref{eq3}), which we restrict this paper to. 
Modelling of skewness and higher-order statistics would, however, require modulation effects to be incorporated.

In real turbulent boundary layers, both self-similar and non-self-similar motions exist and contribute to the individual Reynolds stress components \citep{baars2020,baars2020part2,deshpande2020,yoon2020}.
Therefore, these additional non-self-similar contributions need to be recognized and appropriately accounted for while considering the decomposition in (\ref{eq3}).
They include the fine dissipative scales, as well as those corresponding to the inertial sub-range \citep{perry1986,saddoughi1994}.
These contributions, however, are small relative to those from the inertial motions \citep{perry1986}, and may thus be deemed insignificant for a wall-bounded flow in the limit of $Re_{\tau}$ $\rightarrow$ $\infty$, which the inviscid AEH models.
Other contributions include those from the very-large-scale-motions or superstructures (SS), which are associated with tall and large $\delta$-scaled eddies spanning across the inertial region and contributing substantively to $\overline{u^2}$ and $\overline{v^2}$ \citep{baars2020,baars2020part2,deshpande2020,yoon2020}.
Evidence from the literature suggests that superstructures, however, do not contribute to $\overline{w^2}$, which is confirmed by the wall-scaling exhibited by the 1-D $w$-spectra \citep{bradshaw1967,morrison1992,katul1996,kunkel2006,baidya2017}.
Given the aforementioned characteristics, when considering these superstructures in the context of active and inactive contributions, the motions would also have an inactive signature in $\overline{u^2}$ and $\overline{v^2}$ in the inertial region.
The total inactive contributions can thus be segregated as:
\begin{equation}
\label{eq3a}
\begin{aligned}
\overline{u^2}_{\rm inactive} \; &= \; \overline{{u}^{2}}_{\rm inactive, AE} + \overline{{u}^{2}}_{\rm inactive, SS}, \; \text{and}\\
\overline{v^2}_{\rm inactive} \; &= \; \overline{{v}^{2}}_{\rm inactive, AE} + \overline{{v}^{2}}_{\rm inactive, SS},\\
\end{aligned}
\end{equation}
where $\overline{{u}^{2}}_{\rm inactive, SS}$ and $\overline{{u}^{2}}_{\rm inactive, AE}$ represent inactive contributions from the $\delta$-scaled superstructures and self-similar attached eddies, respectively.
It is the presence of the former, which obscures the pure logarithmic decay of $\overline{{u}^{2}}_{\rm inactive}$ with $z$, as well as the true $k^{-1}_{x}$-scaling in the associated 1-D spectra \citep{jimenez2008,rosenberg2013,baars2020,baars2020part2}.

\subsection{Present contributions}

The present study first proposes a methodology to estimate $\overline{u^{2}}_{\rm active}$ and $\overline{u^{2}}_{\rm inactive}$ in the inertially-dominated region of a canonical wall-bounded flow.
Developing this capability of segregating the active from the inactive component, especially for $u$, is of use to the wall-turbulence modelling community, since it is $u_{\rm active}$ which contributes to the momentum transfer (equation \ref{eq3}).
The present methodology exploits the characteristic of the inactive motions (say at a given wall-normal distance $z_{o}$ in the inertial region) being chiefly created by large eddies relative to the active motions at $z_{o}$; these inactive motions are coherent across a significant wall-normal distance \citep{townsend1976,baars2017}.
For instance, \citet{townsend1961,townsend1976} describes the inactive motions at $z_{o}$ as `swirling' motions that influence the velocity field at all wall heights below $z_{o}$, including the wall-shear stress, via low frequency variations (see also $\S$5.3 in \citet{hwang2015}).
Such motions have their spectral signatures reflected in the $u$-signals recorded at $z_{o}$ and below, down to the wall (say at a reference wall-normal location $z_{r}$). 
Recent work on the 1-D linear coherence spectrum by \citet{baars2017} and \citet{deshpande2019} has shown that a scale-by-scale cross-correlation of the synchronously acquired $u$-signals, at $z_{o}$ and $z_{r}$, isolates the energetic motions coherent across $z_{o}$ and $z_{r}$, which may be deemed as inactive for the case of $z_{r}$ $\ll$ $z_{o}$.
Following (\ref{eq3}), the isolated energy contribution from the inactive motions ($\overline{{u}^{2}}_{\rm inactive}$) can simply be subtracted from the total $u$-energy at $z_{o}$ ($\overline{u^2}$) to yield contributions predominated by the active motions at $z_{o}$.
This makes the present approach different to previous analytical efforts, such as \citet{panton2007}, wherein the active contributions were simply assumed to be proportional to the Reynolds shear stress to estimate the inactive contributions.

The methodology adopted here to segregate the active and inactive contributions, based on direct measurements, is also implemented later to separate the inactive motions into contributions from the self-similar attached eddies ($\overline{{u}^{2}}_{\rm inactive, AE}$) and from the $\delta$-scaled superstructures ($\overline{{u}^{2}}_{\rm inactive, SS}$).
While contributions from the latter are known to be predominant across the inertial region, the self-similar attached eddy contributions to the inactive motions reduce significantly beyond the $\delta$-scaled upper bound of the logarithmic (log) region \citep{baars2020,baars2020part2}.
By choosing the reference wall-normal location at this upper bound, say at a $z_{r}$ $\gg$ $z_{o}$, the scale-by-scale cross-correlation of the synchronously acquired $u$-signals at these $z_{o}$ and $z_{r}$ would isolate $\overline{{u}^{2}}_{\rm inactive, SS}$, which following (\ref{eq3a}) can be used to estimate $\overline{{u}^{2}}_{\rm inactive, AE}$.

To this end, two zero-pressure gradient turbulent boundary layer (ZPG TBL) datasets, comprising multi-point $u$-fluctuations measured synchronously across a wide range of wall-normal (${\Delta}z$ = $\mid{z_{o}}$ - ${z_{r}}\mid$) and spanwise (${\Delta}y$) spacings, are considered.
The datasets include measurements across the inertially-dominated (log) region, and the TBLs span a decade of $Re_{\tau}$, permitting us to test for:
(i) the universal wall-scaling of the $u$-spectra associated with the active motions at $z_{o}$, 
and (ii) the $k^{-1}_{x}$-scaling of the $u$-spectra associated with the self-similar attached eddies inactive with respect to $z_{o}$.
These data are first used to directly compute the 2-D $u$-spectrum \citep{chandran2017,chandran2020}, which gives a map of the energy contributions from eddies of various streamwise (${\lambda}_{x}$ = $2{\pi}/{k_{x}}$) and spanwise (${\lambda}_{y}$ = $2{\pi}/{k_{y}}$) wavelengths coherent across $z_{o}$ and $z_{r}$ \citep{deshpande2020}.
The two-point statistics are then used as an input to a spectral linear stochastic estimation (SLSE; \citet{tinney2006}, \citet{baars2016slse}) based procedure, 
which estimates the subset of the 2-D $u$-energy spectrum at $z_{o}$, associated with specific coherent motions coexisting at $z_{o}$.

\section{ZPG TBL datasets}
\label{data}

Two ZPG TBL datasets, consisting of synchronous multi-point $u$-velocity fluctuations, are considered for analysis in the present study.
One is the $Re_{\tau}$ $\approx$ 2 000 DNS dataset of \citet{sillero2014}, while the other is the $Re_{\tau}$ $\approx$ 14 000 experimental dataset, a part of which has been reported previously in \citet{deshpande2020}. A brief description of the two datasets is presented below.

\begin{table}
\centering
\begin{center}
\begin{tabular}{@{\extracolsep{1pt}} P{1.2cm} P{2.0cm} P{0.8cm} P{1.1cm} P{1.3cm} P{0.35cm} P{1.5cm} P{0.75cm} P{1.25cm} P{1.25cm} }
& \multicolumn{4}{c}{$Re_{\tau}$ $\approx$ 14 000 \citep{deshpande2020}} & & \multicolumn{4}{c}{$Re_{\tau}$ $\approx$ 2 000 \citep{sillero2014}} \\
\cmidrule{2-5} \cmidrule{7-10}
$Set-up$ & $z^{+}_{o}$ & $z^{+}_{r}$ & $T{U_{\infty}}/{\delta}$ & $({\Delta}y)_{max}$ & & $z^{+}_{o}$ & $z^{+}_{r}$ & $({\Delta}x)_{max}$ & $({\Delta}y)_{max}$\\
\vspace{0.15mm}\\
${\Phi}$ & \textbf{15}, 100, 200, 318, 477, 750, 1025, \underline{2250} & $\approx$ $z^{+}_{o}$ & 19 500 & 2.7$\delta$ & & \textbf{15}, 120 -- {\underline{250}} & $=$ $z^{+}_{o}$ & 11.9$\delta$ & 7.6$\delta$\\
\vspace{0.15mm}\\
${\Phi}_{cross}$ & 100, 200, 318, 477, 750, 1025, \underline{2250} & \textbf{15} & 19 500 & 2.5$\delta$ & & 120 -- {\underline{250}} & \textbf{15} & 11.9$\delta$ & 7.6$\delta$\\
\vspace{0.15mm}\\
${\Phi}_{cross}$ & 100, 200, 318 & \underline{2250} & 19 500 & 2.5$\delta$ & & -- & -- & -- & --\\
\hline
\end{tabular}
\caption{A summary of the ZPG TBL datasets comprising synchronized multi-point $u$-signals at $z^{+}_{r}$ and $z^{+}_{o}$ used to compute two types of 2-D $u$-spectra, ${\Phi}$ and ${\Phi}_{cross}$. 
The terminology has been described in $\S$\ref{exp} and figure \ref{fig1}. Underlined values represent the approximate upper bound of the log-region (0.15${Re_{\tau}}$; \citet{marusic2013}), while the values in bold represent the near-wall reference location. Superscript `+' denotes normalization in viscous units.}
\label{tab_exp}
\end{center}
\end{table}

\subsection{Multi-point measurements at $Re_{\tau}$ $\approx$ 14 000}
\label{exp}

\begin{figure}
\begin{center}
\includegraphics[width=1.0\textwidth]{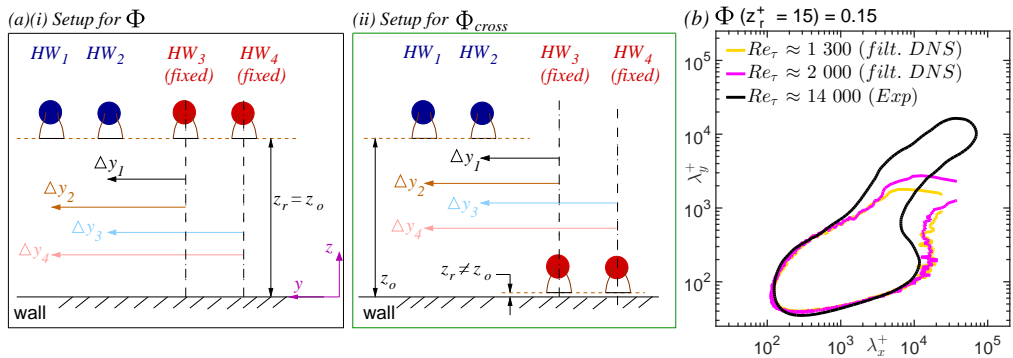}
\caption{(a) Schematic of the experimental set-up in HRNBLWT showing relative positioning and movement of the four hot-wire probes ($HW_{1-4}$) for reconstructing the 2-D correlation corresponding to (i) $\Phi$ and (ii) ${\Phi}_{cross}$. Mean flow direction is along $x$. 
In the case of (ii), $HW_{3-4}$ are positioned at either $z_{r}$ $\ll$ $z_{o}$ or $z_{r}$ $\gg$ $z_{o}$, depending on the desired experiment (table \ref{tab_exp}).
(b) Constant energy contours for ${\Phi}$($z^{+}_{o}$ = $z^{+}_{r}$ $\approx$ 15) = 0.15, computed from the present experimental and the converged DNS dataset of \citet{sillero2014}, plotted as a function of viscous-scaled wavelengths. Estimates from the DNS are box-filtered along $y$ to mimic the spatial resolution of the hotwire sensors.}
\label{fig1}
\end{center}
\end{figure}

The high $Re_{\tau}$ dataset was acquired in the large Melbourne wind tunnel (HRNBLWT) under nominal ZPG conditions and low free-stream turbulence levels \citep{marusic2015} across its working section dimensions of $\simeq$ 0.92\;m $\times$ 1.89\;m $\times$ 27\;m. 
The very long length (27\;m), and capability to generate free-stream speeds of up to 45\;ms$^{-1}$, permit ZPG TBL measurements to the order of $Re_{\tau}$ $~(=\delta U_{\tau}/\nu)$ $\approx$ 26 000 in this facility. 
In the present study, all measurements were conducted at a location approximately 20\;m from the start of the working section, at a free-stream speed of $U_{\infty}$ $\approx$ 20\;$\text{ms}^{-1}$, resulting in a ZPG TBL at $Re_{\tau}$ $\approx$ 14 000.
The TBL thickness $\delta$ here is estimated via the modified Coles law of the wake fit \citep{jones2001} for both datasets.
The multi-point measurements were made possible by a unique experimental set-up \citep{chandran2017} employing four hotwire probes ($HW_{1-4}$), the arrangement of which is depicted in figure \ref{fig1}(a).
Wollaston hotwire probes of diameter, $d$ $\approx$ 2.5\;$\mu$m and exposed sensor length, $l$ $\approx$ 0.5\;mm were used for all the measurements, resulting in an acceptable length-to-diameter ratio of approximately 200 \citep{hutchins2009} and a viscous-scaled sampling length, ${l^{+}} ~(={l}{U_{\tau}}/{\nu})$ $\approx$ 22 for the given measurements. 
This hotwire length is sufficiently small compared to the energetic spanwise wavelengths in the inertial region, which can be inferred from the spanwise spectra of the $u$-velocity component from any published DNS dataset (for instance, see figure 9 of \citet{mklee2015}). 
The sensors were operated in a constant temperature mode using an in-house Melbourne University Constant Temperature Anemometer (MUCTA) at an overheat ratio of 1.8 and at a viscous-scaled sampling rate, ${\Delta}T^{+}$ $\equiv$ ${U^{2}_{\tau}}/{({\nu}{f_{s}})}$ $\approx$ 0.5, where $f_{s}$ refers to sampling frequency.

The experimental set-up, as depicted in figure 1, allows $HW_{1-2}$ to be traversed in the spanwise direction at a consistent wall-normal distance of $z_{o}$, while $HW_{3-4}$ remain stationary at a fixed spanwise and wall-normal ($z_{r}$) location throughout the measurement.
To calibrate the probes, the same procedure as that employed by \citet{chandran2017} was implemented with $HW_{1}$, $HW_{2}$ and $HW_{4}$ simultaneously calibrated at a common wall-normal location by using the free-stream calibrated $HW_{3}$ as a reference.
Simultaneously acquired $u$-signals from the four hotwires are used to reconstruct the two-point correlation:
\begin{equation}
\label{eq4}
\begin{aligned}
{R_{{u_{o}}{u_{r}}}}({z_{o}},{z_{r}};{\Delta}x,{\Delta}y) = \overline{ {{u}({z_{r}};x,y)}{{u}({z_{o}};x+{\Delta}x,y+{\Delta}y)} }
\end{aligned}
\end{equation}
for the ${\Delta}y$ range, 0 $\le$ ${\Delta}y$ $\le$ $({\Delta}{y})_{max}$ and the total sampling duration ($T$) of the $u$-signals listed in table \ref{tab_exp},
with the overbar denoting ensemble time average.
Taylor's frozen turbulence hypothesis, which considers all the coherent structures coexisting at $z_{o}$ to be convecting at the mean velocity at $z_{o}$ (i.e. $U_{c}$ = $U(z_{o})$), is used to convert ${R_{{u_{o}}{u_{r}}}}$ from a function of time to that of ${\Delta}{x}$, with $U_{c}$ denoting the convection velocity assumed at $z_{o}$.
Following this, the 2-D Fourier transform of ${R_{{u_{o}}{u_{r}}}}$ is computed to obtain the 2-D spectrum as:
\begin{equation}
\label{eq5}
\begin{split}
{{\phi}_{{u_{o}}{u_{r}}}}({z_{o}},{z_{r}};{k_{x}},{k_{y}}) = \int \int_{-{\infty}}^{\infty} {R_{{u_{o}}{u_{r}}}}({z_{o}},{z_{r}};{\Delta}x,{\Delta}y) {e^{-{j2{\pi}({k_{x}}{\Delta}x + {k_{y}}{\Delta}y)}}} d({\Delta}x)d({\Delta}y), 
\end{split}
\end{equation}    
with $j$ a unit imaginary number.

For this study, we are only concerned with two types of 2-D spectra, ${\Phi}$ and ${\Phi}_{cross}$ which are defined as: 
\begin{equation}
\label{eq6}
\begin{split}
{\Phi}(z^{+}_{o};{{\lambda}_{x}},{{\lambda}_{y}}) &= {\mid}{k^{+}_{x}}{k^{+}_{y}}{{\phi}^{+}_{{u_{o}}{u_{o}}}}({z^{+}_{o}};{{\lambda}_{x}},{{\lambda}_{y}}){\mid} \; \text{and}\\
{{\Phi}_{cross}}(z^{+}_{o},z^{+}_{r};{{\lambda}_{x}},{{\lambda}_{y}}) &= {\mid}{k^{+}_{x}}{k^{+}_{y}}{{\phi}^{+}_{{u_{o}}{u_{r}}}}({z^{+}_{o}},{z^{+}_{r}};{{\lambda}_{x}},{{\lambda}_{y}}){\mid},
\end{split}
\end{equation}  
with the ${R_{{u_{o}}{u_{r}}}}$ corresponding to the former and latter, reconstructed via hotwire arrangements depicted in figure \ref{fig1}(a,i) and \ref{fig1}(a,ii), respectively.
Here, $z^{+}_{o}$ = $\frac{{z_{o}}{U_{\tau}}}{\nu}$ and $k^{+}_{x}$ = $\frac{{k_{x}}{\nu}}{U_{\tau}}$ (with similar definitions for other associated terms), where the superscript `+' indicates normalization in viscous units.
Table \ref{tab_exp} details the exact wall-normal locations for which ${\Phi}$ and ${\Phi}_{cross}$ are computed, with (${\mid}{\mid}$) referring to the modulus operation.
The present analysis is focused in the inertially-dominated region, considered nominally to exist beyond $z^{+}_{o}$ $\gtrsim$ 100 \citep{nickels2005,mklee2015,chandran2017,baars2020}, based on the empirical evidence discussed in $\S$\ref{intro}. 
While ${\Phi}$ represents contributions from all coexisting motions at $z_{o}$, ${\Phi}_{cross}$ consists of contributions from only those motions that are coherent across $z_{o}$ and $z_{r}$ \citep{deshpande2020}.
Both these spectra are used as an input to the SLSE methodology ($\S$\ref{slse}) to estimate subsets of ${\Phi}$($z_{o}$) representing contributions from a specific family of coherent motions coexisting at $z_{o}$.
${\Phi}_{cross}$ has been estimated for two different reference wall-normal positions ($z_{r}$; table \ref{tab_exp}), each targeted at isolating specific contributions.
The measurements to obtain ${\Phi}_{cross}$($z^{+}_{o},z^{+}_{r}$ $\approx$ 0.15$Re_{\tau}$), however, were conducted following the same methodology as that adopted for ${\Phi}_{cross}$($z^{+}_{o},z^{+}_{r}$ $\approx$ 15), which have been reported previously in \citet{deshpande2020} and may be consulted for further details.

The present study also reports the first measurements of ${\Phi}$ in the near-wall region ($z^{+}_{o}=z^{+}_{r}$ $\approx$ 15), which is required as per the SLSE methodology ($\S$\ref{slse}) being adopted in the present study.
Figure \ref{fig1}(b) compares the constant energy contour for this experimentally estimated ${\Phi}$ against the same computed from the converged 2-D $u$-correlations available from the DNS dataset of \citet{sillero2014}.
While a reasonable overlap of contours is observed in the small-scale range (figure \ref{fig1}(b)), when plotted as a function of viscous-scaled wavelengths, a prominent `footprint' can be noted appearing for the large scales with increase in $Re_{\tau}$.
This is representative of the increasing influence of the large scales in the near-wall region with increase in $Re_{\tau}$, as discussed by \citet{hutchins2007} and \citet{hutchins2009}.
Here, the spectra from the DNS are box-filtered for better one-to-one comparison with the experimental spectrum, wherein the energy in the small-scales is underestimated due to the spatial resolution of the hotwire sensor \citep{hutchins2009}.
The box-filtering is carried out along the $y$-direction, by following the same methodology as outlined in \citet{chin2009}, taking into consideration the viscous-scaled hotwire sensor length corresponding to the measurements ($l^{+}$ $\approx$ 22).
Another thing to note here is that the contour corresponding to the experimental spectrum deviates significantly from the low $Re_{\tau}$ DNS estimates at large wavelengths.
This is possibly due to the failure of Taylor's hypothesis for these large-scales in the near-wall region \citep{del2009,monty2009dns}.
This inconsistency, however, doesn't affect any of the forthcoming analysis since all the calculations ($\S$\ref{slse}) for the experimental dataset are carried out in the frequency domain before converting to ${\lambda}_{x}$ via Taylor's hypothesis.

\subsection{DNS dataset}
\label{dns}

A low $Re_{\tau}$ dataset from the ZPG TBL DNS of \citet{sillero2014} is also considered in the present study.
Thirteen raw DNS volumes, each of which is a subset of their full computational domain between $x$ $\approx$ 28.4$\delta$ and $x$ $\approx$ 40.3$\delta$, are selected to ensure a limited $Re_{\tau}$ increase along $x$.
Streamwise velocities $u$($z^{+}_{o}$;$x$,$y$) extracted from these fields are used to compute ${\Phi}$($z^{+}_{o}$) and ${\Phi}_{cross}$($z^{+}_{o},z^{+}_{r}$ $\approx$ 15)  following (\ref{eq4}) -- (\ref{eq6}), at $z^{+}_{o}$ and $z^{+}_{r}$ consistent with the experimental dataset (table \ref{tab_exp}). 
A similar analysis is also conducted using the instantaneous wall-normal velocity fluctuations, $w$($z^{+}_{o}$;$x$,$y$) extracted from this dataset.
It is used to establish the efficacy of the SLSE-based methodology being implemented here to segregate active and inactive contributions, the results from which are discussed in appendix 1.
 
\section{Energy decomposition into active and inactive contributions}
\label{slse}

As discussed in $\S$\ref{intro}, the inactive motions at $z_{o}$ are predominantly large motions (with respect to $z_{o}$) that are coherent across a significant wall-normal distance. This forms the basis for decomposing ${\Phi}$($z_{o}$).
Classically, the size and scaling of the coherent structures have been interpreted via two-point cross-correlations \citep{ganapathisubramani2005,hutchins2007}.
Correlations represent contributions from a wide range of scales which, during the ensemble averaging procedure, do not distinguish the individual contributions from the small and large motions \citep{baars2017,deshpande2019}.
Therefore, the present investigation has been conducted entirely in the spectral domain.
Previous studies \citep{balakumar2007} employing the spectral approach have utilized a sharp streamwise spectral cut-off to segregate the large motions from the rest of the turbulence, which inherently comes with a drawback that the estimates are cut-off dependent.

\begin{figure}
\begin{center}
\includegraphics[width=1.0\textwidth]{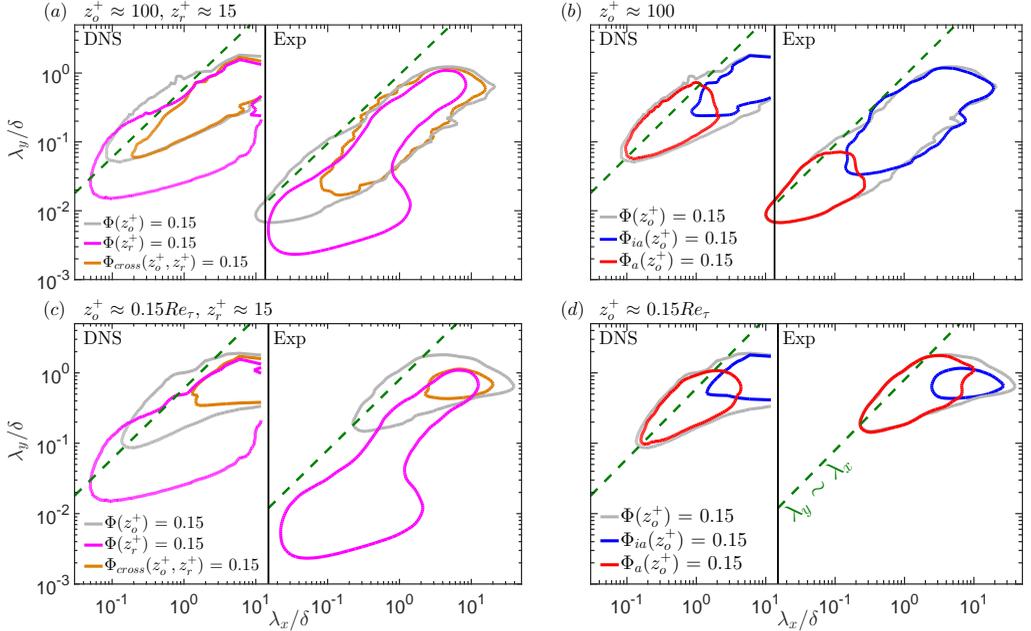}
\caption{(a,c) Constant energy contours for ${\Phi}$($z^{+}_{o}$), ${\Phi}_{cross}$($z^{+}_{o}$, $z^{+}_{r}$ $\approx$ 15) and ${\Phi}$($z^{+}_{r}$ $\approx$ 15) at energy level of 0.15 plotted for (a) $z^{+}_{o}$ $\approx$ 100 and (c) $z^{+}_{o}$ $\approx$ 0.15$Re_{\tau}$. 
(b,d) Constant energy contours for ${\Phi}_{ia}$($z^{+}_{o}$) and ${\Phi}_{a}$($z^{+}_{o}$), computed via (\ref{eq7}) and (\ref{eq8}), plotted at the same energy level and $z^{+}_{o}$ as in (a,c), respectively. 
In (a-d), contours on the left side correspond to those computed from the DNS data while those on the right are from the experimental data.
Dashed green lines represent the linear relationship, ${\lambda}_{y}$ $\sim$ ${\lambda}_{x}$.}
\label{fig2}
\end{center}
\end{figure} 

Here, the intention is to perform an unconditional linear decomposition of ${\Phi}$($z_{o}$) into its inactive and residual component (figure \ref{fig2}) by utilizing the scale-by-scale coupling between $u$-signals simultaneously measured at two wall-normal locations, $z^{+}_{o}$ (in the inertially-dominated region) and ${z^{+}_{r}}$ $\approx$ 15, ensuring $z^{+}_{r}$ $\ll$ $z^{+}_{o}$.
A linear decomposition was deemed sufficient for this purpose given the fact that the coupling has been computed between velocity signals at both ends \citep{guezennec1989,baars2016slse}, and that the present interests are limited to the second-order velocity statistics ($\S$\ref{intro}).
${\Phi}_{cross}$($z^{+}_{o},z^{+}_{r}$ $\approx$ 15), which is considered here at various $z^{+}_{o}$ for both the experimental and DNS datasets, represents this scale-by-scale coupling.  
On comparing ${\Phi}_{cross}$($z^{+}_{o},z^{+}_{r}$ $\approx$ 15) and ${\Phi}$($z^{+}_{o}$) contours from the two datasets at various $z^{+}_{o}$ in figures \ref{fig2}(a,c), 
the former is representative of energetic large-scales that can be associated with the motions inactive at $z^{+}_{o}$.
It is evident that ${\Phi}_{cross}$($z^{+}_{o},z^{+}_{r}$ $\approx$ 15) also inherently comprises energy contributions from the $\delta$-scaled superstructures (${\lambda}_{x}$ $\gtrsim$ 6$\delta$), which are known to extend from the wall and span across the inertial region \citep{baars2020,baars2020part2,deshpande2020,yoon2020}.
We use ${\Phi}_{cross}$ in conjunction with the SLSE \citep{tinney2006,baars2016slse} to obtain a linear stochastic estimate of the spectrum (${\Phi}_{ia}$) associated with the inactive motions at $z_{o}$ following: 
\begin{equation}
\label{eq7}
{{{\Phi}_{ia}}(z^{+}_{o};{\lambda_{x}},{\lambda_{y}})} = { \frac{ {\left[{{\Phi}_{cross}}({z^{+}_{o}}, {z^{+}_{r}} \approx 15;{\lambda_{x}},{\lambda_{y}}) \right]}^{2} }{ {{\Phi}}({z^{+}_{r}} \approx 15;{\lambda_{x}},{\lambda_{y}}) } }.
\end{equation} 
Interested readers may refer to appendix 1 to see the step-by-step procedure to arrive at the expression in (\ref{eq7}).
The mathematical operation in the above equation suggests ${{\Phi}_{ia}}$($z^{+}_{o}$) to be essentially a normalized version of ${{\Phi}_{cross}}$(${z^{+}_{o}},{z^{+}_{r}}$ $\approx$ 15), with the scale-by-scale normalization done by ${{\Phi}}$(${z^{+}_{r}}$ $\approx$ 15), the contours for which are also plotted in figures \ref{fig2}(a,c).
It should be noted here that the calculations in (\ref{eq7}) are carried out in the frequency domain for the experimental dataset, with the conversion to ${\lambda}_{x}$ by invoking Taylor's hypothesis, using $U_{c}$ $=$ $U$($z_{o}$) \citep{baars2016slse,baars2017}.
Following the linear superposition assumption in (\ref{eq3}), ${\Phi}_{ia}$ can be simply subtracted from ${\Phi}$ to leave a residual:
\begin{equation}
\label{eq8}
{{{\Phi}_{a}}(z^{+}_{o};{\lambda_{x}},{\lambda_{y}})} = {{\Phi}(z^{+}_{o};{\lambda_{x}},{\lambda_{y}})} - {{{\Phi}_{ia}}(z^{+}_{o};{\lambda_{x}},{\lambda_{y}})},
\end{equation}
with ${\Phi}$, ${\Phi}_{ia}$ and ${\Phi}_{a}$ representative of $\overline{u^2}^{+}$, $\overline{u^{2}}^{+}_{\rm inactive}$ and $\overline{u^{2}}^{+}_{\rm active}$, respectively.
If the flow consisted of only active and inactive inertial motions, ${\Phi}_{a}$ and ${\Phi}_{ia}$ would be the active and inactive component, respectively.
However, we refer to ${\Phi}_{a}$ as the residual spectrum, given that it also comprises small contributions from the fine dissipative scales as well as those corresponding to the inertial sub-range ($\S$\ref{intro}).
We limit their influence in the present analysis by focusing our investigation on the high energy contours of ${\Phi}_{a}$, which are associated predominantly with the inertial active motions. 

Figures \ref{fig2}(b,d) show the constant energy contours for the two components ${\Phi}_{ia}$($z^{+}_{o}$) and ${\Phi}_{a}$($z^{+}_{o}$), computed via (\ref{eq7}) and (\ref{eq8}), using the corresponding inputs plotted in figures \ref{fig2}(a,c), respectively.
While ${\Phi}_{ia}$ takes up the large-scale portion of ${\Phi}$, ${\Phi}_{a}$ is restricted to the small-scale end of the spectrum.
This is in spite of the fact that ${\Phi}_{cross}$($z^{+}_{o},z^{+}_{r}$ $\approx$ 15) also comprises contributions from relatively small scales at $z^{+}_{o}$ $\approx$ 100 (figure \ref{fig2}(a)) and can be explained by the linear transfer kernel (equations \ref{app_eq12} and \ref{app_eq13}), which has been computed at various $z^{+}_{o}$ for the DNS dataset and shown in figure \ref{fig1_app}(a) in appendix 1.
Interestingly, at $z^{+}_{o}$ $\approx$ 100 (figure \ref{fig2}(b)), both ${\Phi}_{a}$($z^{+}_{o}$) and ${\Phi}_{ia}$($z^{+}_{o}$) can be seen to follow the ${\lambda}_{y}$ $\sim$ ${\lambda}_{x}$ relationship representative of geometric self-similarity, which is otherwise obscured for ${\Phi}$ in the intermediate and large-scale range \citep{chandran2017,chandran2020,deshpande2020}.
The self-similar characteristic of ${\Phi}_{ia}$ and ${\Phi}_{a}$ is consistent with the hypothesis of \citet{townsend1961,townsend1976}, who originally described both the active and inactive motions to be associated purely with the attached eddy contributions, but conforming to a different range of scales: 
the active motions at $z^{+}_{o}$ conform to the attached eddies with height, $\mathcal{H}$ $\sim$ ${\mathcal{O}}(z_{o})$, 
while the inactive motions conform to relatively large eddies with ${\mathcal{O}}(z_{o})$ $\ll$ $\mathcal{H}$ $\lesssim$ ${\mathcal{O}}(\delta)$ (see $\S$\ref{intro}).
Consequently, the contribution from the attached eddies to ${\Phi}_{ia}$ reduces with increase in $z^{+}_{o}$, with energy contours at $z^{+}_{o}$ $\approx$ 0.15$Re_{\tau}$ (figure \ref{fig2}(d)) corresponding predominantly to the tall $\delta$-scaled superstructures coexisting across the inertial region.
This likely explains why ${\Phi}_{ia}$ contours do not align along ${\lambda}_{y}$ $\sim$ ${\lambda}_{x}$ at $z^{+}_{o}$ farthest from the wall.
It also forms the basis for choosing $z^{+}_{r}$ $\approx$ 0.15$Re_{\tau}$ as a reference wall height while implementing the SLSE methodology to isolate the superstructure contributions, which will be discussed later in $\S$\ref{inactive_AE}.
${\Phi}_{a}$, on the other hand, comprises a significant range of scales irrespective of the change in $z^{+}_{o}$, with the contours simply shifting to relatively larger scales, which is suggestive of its distance-from-the-wall ($z_{o}$) scaling.
Having defined the procedure to obtain ${\Phi}_{a}$ and ${\Phi}_{ia}$, next we test for $z_{o}$- and $\delta$-scaling to verify the extent to which the respective spectra can be associated with the active and inactive motions.

\begin{figure}
\begin{center}
\includegraphics[width=0.9\textwidth]{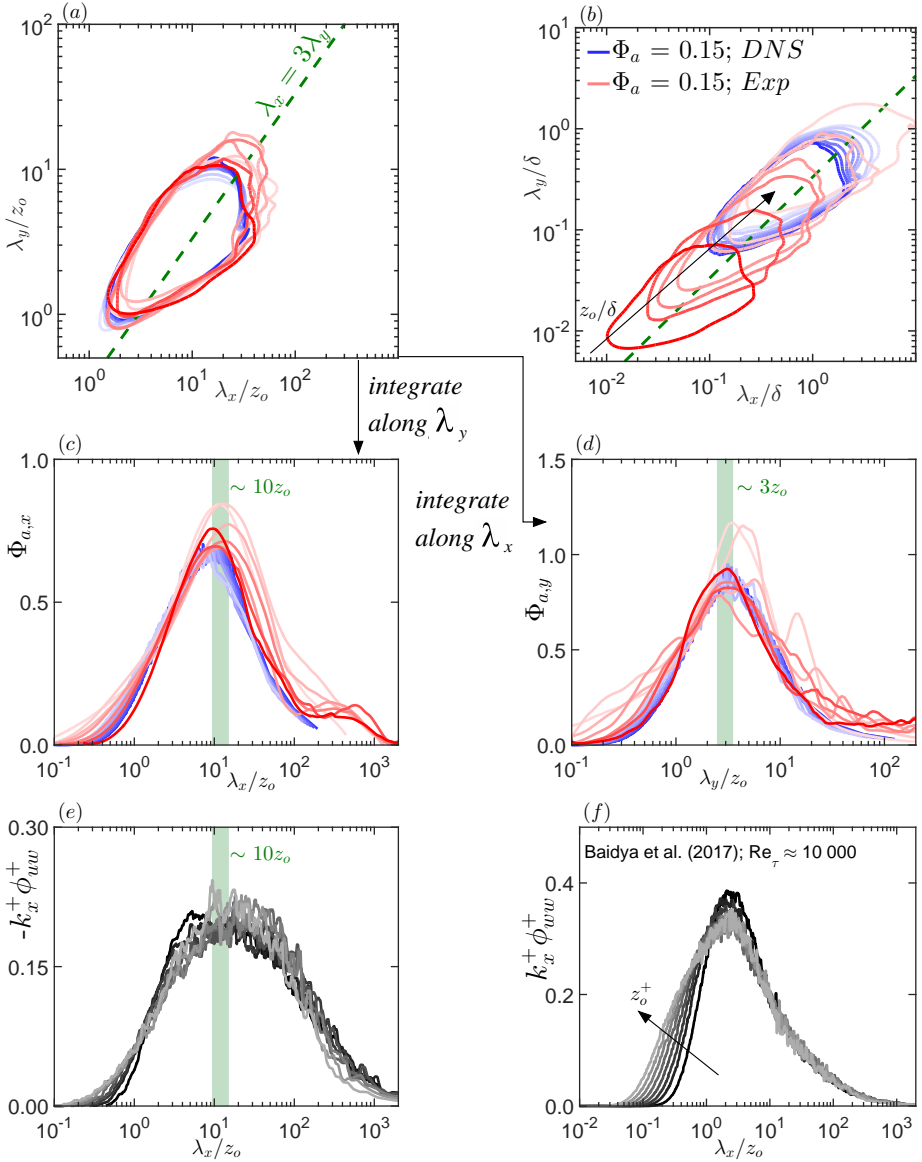}
\caption{(a,b) Constant energy contours for ${\Phi}_{a}$($z^{+}_{o}$) at energy level of 0.15 plotted for various $z^{+}_{o}$ as a function of wavelengths scaled with (a) $z_{o}$ and (b) $\delta$. 
Contours in red and blue correspond to ${\Phi}_{a}$ estimated for the experimental and DNS datasets respectively (table \ref{tab_exp}), with dark to light shading indicating an increase in $z^{+}_{o}$ following 100 $\lesssim$ $z^{+}_{o}$ $\lesssim$ 0.15${Re_{\tau}}$ for $Re_{\tau}$ corresponding to respective datasets. 
Dashed green lines represent the linear relationship, ${\lambda}_{x}$ = 3${\lambda}_{y}$.
(c,d) ${\Phi}_{a}$($z^{+}_{o}$) integrated across ${\lambda}_{y}$ and ${\lambda}_{x}$ to obtain its corresponding 1-D version as a function of (c) ${\lambda}_{x}$ and (d) ${\lambda}_{y}$ respectively, each plotted with wavelengths scaled by $z_{o}$. Same colour coding is followed as that described for (a,b).
(e,f) Pre-multiplied streamwise 1-D cospectra/spectra for the (e) Reynolds shear stress and (f) wall-normal velocity plotted as a function of ${\lambda}_{x}$ scaled with $z_{o}$. This data is from the $Re_{\tau}$ $\approx$ 10 000 dataset of \citet{baidya2017} for various $z^{+}_{o}$. 
Dark to light shading corresponds to the increase in $z^{+}_{o}$ following 100 $\lesssim$ $z^{+}_{o}$ $\lesssim$ 0.15${Re_{\tau}}$, where $Re_{\tau}$ $\approx$ 10 000.}
\label{fig3}
\end{center}
\end{figure} 

\section{Active component of the streamwise velocity spectrum}
\label{active}

\begin{figure}
\begin{center}
\includegraphics[width=0.9\textwidth]{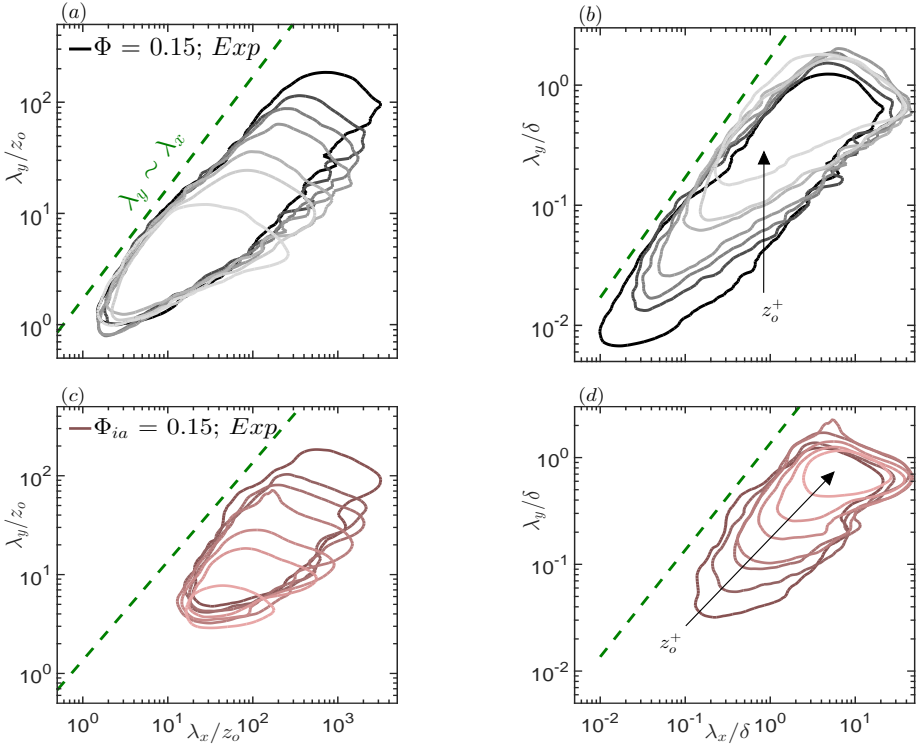}
\caption{Constant energy contours for (a,b) ${\Phi}$($z^{+}_{o}$) and (c,d) ${\Phi}_{ia}$($z^{+}_{o}$) at energy level of 0.15 plotted for various $z^{+}_{o}$ as a function of wavelengths scaled with (a,c) $z_{o}$ and (b,d) $\delta$, respectively. All data in (a-d) corresponds to the high $Re_{\tau}$ experimental dataset reported in table \ref{tab_exp}, with dark to light shading indicating an increase in $z^{+}_{o}$ following 100 $\lesssim$ $z^{+}_{o}$ $\lesssim$ 0.15${Re_{\tau}}$. Dashed green lines represent the linear relationship, ${\lambda}_{y}$ $\sim$ ${\lambda}_{x}$.}
\label{fig4}
\end{center}
\end{figure}

Figures \ref{fig3}(a,b) show the constant energy contours of ${\Phi}_{a}$ (= 0.15), computed from both DNS and experimental datasets, plotted as a function of wavelengths scaled with $z_{o}$ and ${\delta}$, respectively.
The contours are plotted for ${\Phi}_{a}$ across 100 $\lesssim$ $z^{+}_{o}$ $\lesssim$ 0.15$Re_{\tau}$ and are seen to reasonably follow wall-scaling, that is, when the wavelengths are normalized by $z_{o}$. In contrast, no such collapse is observed when the wavelengths are scaled with $\delta$. 
It is noted that this behaviour is only apparent after separating ${\Phi}_{a}$ from ${\Phi}$. 
For comparison, figure \ref{fig4} in $\S$\ref{inactive} shows the corresponding ${\Phi}$ results which exhibit both $z_{o}$- and $\delta$-scaling in the intermediate and large-scale wavelength ranges, respectively (due to the wall-parallel velocity field associated with both the active and inactive motions \citep{bradshaw1967,baidya2017}).

The $z_{o}$-scaling behaviour noted for ${\Phi}_{a}$ is consistent with active motions. This can be seen by comparing the scaling behaviour of 1D $w$-spectra and 1D $uw$-cospectra, which have been shown to follow wall-scaling and exhibit a behaviour exclusively associated with active motions \citep{bradshaw1967,morrison1992,baidya2017}.
To this end, ${\Phi}_{a}$ is integrated along ${\lambda}_{y}$ and ${\lambda}_{x}$ to obtain the corresponding premultiplied 1-D spectra as a function of ${\lambda}_{x}$ (${\Phi}_{a,x}$; figure \ref{fig3}(c)) and ${\lambda}_{y}$ (${\Phi}_{a,y}$; figure \ref{fig3}(d)), respectively.
Also plotted, are the pre-multiplied 1-D $w$-spectra (figure \ref{fig3}(f)) and $uw$-cospectra (figure \ref{fig3}(e)) at 100 $\lesssim$ $z^{+}_{o}$ $\lesssim$ 0.15$Re_{\tau}$ from the $Re_{\tau}$ $\approx$ 10 000 dataset of \citet{baidya2017}, measured at the same experimental facility as \citet{deshpande2020}.
When the wavelengths are scaled with $z_{o}$, the 1-D spectra in figures \ref{fig3}(c-f) are observed to collapse for ${\lambda}$ $\gtrsim$ ${z_{o}}$, in-line with the characteristics of active motions.
Further, both ${\Phi}_{a,x}$ and ${k^{+}_{x}}{{\phi}^{+}_{uw}}$ peak at ${\lambda}_{x}$ $\sim$ 10$z_{o}$, supporting the argument that the motions associated with ${\Phi}_{a}$ contribute to the Reynolds shear stress and can hence be deemed active in the sense of \citet{townsend1961,townsend1976}.
The efficacy of the present SLSE-based methodology, to extract energetic contributions from the active motions, can also be tested by implementing it on similar two-point statistics computed for the $w$-velocity component. 
Given that the $w$-component is associated exclusively with the active motions ($\S$\ref{intro}), the present methodology can be deemed effective if it reveals negligible energy contributions from the inactive spectrum for the $w$-component. Interested readers may refer to appendix 1 where the SLSE analysis conducted on the $w$-component has been discussed.

Small scales (${\lambda}$ $\ll$ ${z_{o}}$), which correspond to the viscous dissipative scales or those following the inertial sub-range scaling, do not scale with distance from the wall, explaining the deviation from the collapse of the 1-D spectra in figures 3(c,d,f).
A similar deviation, although at a much smaller magnitude, is also observed for the Reynolds shear stress cospectra, which eventually drops to zero at ${{\lambda}_{x}}/{z_{o}}$ $\lesssim$ 0.2 owing to the approximate isotropy of these fine scales \citep{saddoughi1994}.
On a side note, the reasonable agreement between ${\Phi}_{a,x}$ estimated from the DNS and experimental datasets also validates the use of the local mean velocity as the convection velocity ($U_{c}$ = $U(z_{o})$) for the active motions, which seems intuitive given these are localized at $z_{o}$.

The present analysis, which is conducted along both the $x$ and $y$ directions, also reveals the dominant spanwise wavelength corresponding to the active motions, i.e. ${\lambda}_{y}$ $\sim$ 3${z_{o}}$ (figure \ref{fig3}(d)).
This yields the dominant streamwise/spanwise aspect ratio of ${\lambda}_{x}$/${\lambda}_{y}$ $\sim$ 3 for these motions, which is found to be true across a decade of $Re_{\tau}$ (indicated by dashed line in figures \ref{fig3}(a,b)).
A similar SLSE-based analysis, as implemented here for the $u$-velocity spectrum, was conducted on the DNS dataset to analyze the active component of the $v$- and $w$-velocity spectrum (not shown here for brevity).
These components were also found to exhibit wall-scaling, across the inertially-dominated region, with the contours of the spectrum following the self-similar relationship, ${\lambda}_{x}$/${\lambda}_{y}$ $\sim$ 1 and ${\lambda}_{x}$/${\lambda}_{y}$ $\sim$ 1.4 for the $v$ and $w$-velocity spectrum, respectively.
Interestingly, the aspect ratio found for the active $u$-spectrum matches that of the self-similar wall-coherent vortex clusters (${\lambda}_{x}$ $\sim$ 2-3${\lambda}_{y}$) investigated by \citet{del2006} and \citet{hwang2015}, revealing information which may be useful for modelling the active motions in future works.
The close agreement with \citet{hwang2015} further suggests ${\Phi}_{a}$ and ${\Phi}_{ia}$, both of which comprise of prominent self-similar contributions (figures \ref{fig2},\ref{fig3},\ref{fig4}), correspond well with the two component attached eddy structure proposed by \citet{hwang2015} for a wall-bounded turbulent flow.
In their case, \citet{hwang2015} defined motions at a given spanwise scale to be composed of two distinct components: the first is the long streaky flow structure, attached to the wall and having significant turbulent kinetic energy, but inactive in the inner-region. 
The energy contributions from these motions are represented by ${\Phi}_{ia}$.
While, the second component corresponds to the short and tall self-similar vortex packets which are active in the inner-region, and hence would contribute to ${\Phi}_{a}$.

\begin{figure}
\begin{center}
\includegraphics[width=0.9\textwidth]{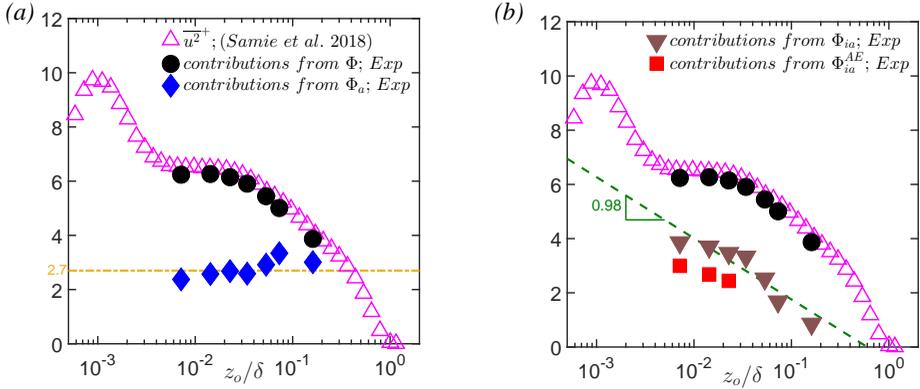}
\caption{Comparison of the normalized streamwise turbulence intensities obtained by integrating ${\Phi}$ (= ${\overline{u^2}}^{+}$), ${\Phi}_{a}$ ($\approx$ $\overline{u^{2}}^{+}_{\rm active}$), ${\Phi}_{ia}$ ($\approx$ $\overline{u^{2}}^{+}_{\rm inactive}$) and ${\Phi}^{AE}_{ia}$ ($\approx$ $\overline{u^{2}}^{+}_{\rm inactive, AE}$) for the high $Re_{\tau}$ experimental dataset described in table \ref{tab_exp}. 
Also plotted for comparison is the well-resolved ${\overline{u^2}}^{+}$ measured by \citet{samie2018} across a ZPG TBL maintained at an $Re_{\tau}$ comparable to the present experimental dataset. 
The dashed green line in (b) represents the logarithmic decay of ${\overline{u^2}}^{+}$ described by (\ref{eq1}) with $A_{1}$ $=$ 0.98 \citep{baars2020part2}, while the dash-dotted golden line in (a) represents a constant ${\overline{u^2}}^{+}$ = 2.7.}
\label{fig5}
\end{center}
\end{figure} 

\section{Inactive component of the streamwise velocity spectrum}
\label{inactive}

Figure \ref{fig4} shows the constant energy contours of ${\Phi}$ (figures \ref{fig4}(a,b)) and ${\Phi}_{ia}$ (figures \ref{fig4}(c,d)), computed for the experimental dataset, plotted as a function of wavelengths scaled with $z_{o}$ (figures \ref{fig4}(a,c)) and $\delta$ (figures \ref{fig4}(b,d)).
These contours are plotted at the same energy level and for the same $z^{+}_{o}$, as in figures \ref{fig3}(a,b).
Consistent with the observations of \citet{bradshaw1967} and \citet{baidya2017} for the 1-D $u$-spectra, ${\Phi}$ contours can be observed to be exhibiting $z_{o}$-scaling in the intermediate scales ($\mathcal{O}$(1) $\lesssim$ ${\lambda}/{z_{o}}$ $\lesssim$ $\mathcal{O}$(10)) and $\delta$-scaling for the large scales (${\lambda}$ $\gtrsim$ $\mathcal{O}$($\delta$)).
This is due to contributions from both the active as well as the inactive motions to ${\Phi}$.

${\Phi}_{ia}$ also exhibits both $z_{o}$- and ${\delta}$-scaling, with the scale range for $z_{o}$-scaling, however, much narrower than that observed for ${\Phi}$.
This can be attributed to the fact that ${\Phi}_{ia}$($z_{o}$) comprises contributions from the attached eddies of height, ${\mathcal{O}}$($z_{o}$) $\ll$ ${\mathcal{H}}$ $\lesssim$ ${\mathcal{O}}$($\delta$), as well as the $\delta$-scaled superstructures ($\S$\ref{intro},$\S$\ref{slse}). 
It means that the attached eddy contributions form a considerable portion of the total inactive contributions at any $z_{o}$ close to the wall, due to which a clear ${\lambda}_{y}$ $\sim$ ${\lambda}_{x}$ trend is discernible in ${\Phi}_{ia}$($z_{o}$). 
\citet{townsend1976}, however, described the classification of an eddy as `active' or `inactive' to be a relative concept, dependent on the wall-normal location under consideration. 
Hence, the tall attached eddies which are inactive relative to $z^{+}_{o}$ $\approx$ 100 may qualify as active at greater wall-heights. 
This explains the narrowing down of the ${\Phi}_{ia}$($z_{o}$) contours to the largest scales with increase in $z^{+}_{o}$ (figure \ref{fig4}(d)), until only the superstructure contributions remain at $z^{+}_{o}$ $\approx$ 0.15$Re_{\tau}$.
The latter explains the deviation of the contours from the linear relationship,  as $z^{+}_{o}$ moves away from the wall.

The reduction in the attached eddy contributions, with increase in $z^{+}_{o}$, translates into a drop of the cumulative streamwise turbulence intensity, i.e. $\iint_{0}^{\infty} {{\Phi}_{ia}} {{d(ln{{\lambda}_{x}})} {d(ln{{\lambda}_{y}})}}$, plotted in figure \ref{fig5}(b) for the experimental dataset. 
Also shown alongside in figure \ref{fig5}(a) are cumulative contributions obtained by integrating ${\Phi}$ and ${\Phi}_{a}$ for 100 $\lesssim$ $z^{+}_{o}$ $\lesssim$ 0.15$Re_{\tau}$.
Figure \ref{fig5} also includes, for reference, the well-resolved ${\overline{u^2}}^{+}$ profile of \citet{samie2018} across the entire boundary layer, as well as a log law with $A_{1}$ $=$ 0.98 proposed by \citep{baars2020part2}.
As is evident from the plot, the contributions from both ${\Phi}$ and ${\Phi}_{ia}$ decay with $z/{\delta}$ very similarly, however, they only approximately follow the $A_{1}$ $=$ 0.98 log law.
This disagreement can be associated with the $\delta$-scaled superstructure contributions existing in both $\Phi$ and ${\Phi}_{ia}$ \citep{jimenez2008,baars2020part2}, given that the expressions in (\ref{eq1}) are valid strictly for self-similar attached eddy contributions alone ($\S$\ref{intro}).
An attempt is thus made to remove this superstructure contribution from ${\Phi}_{ia}$ in the next sub-section, by following the same SLSE-based methodology discussed previously in $\S$\ref{slse}.
Returning to figure \ref{fig5}, a similar variation for both the profiles obtained on integrating $\Phi$ and ${\Phi}_{ia}$ leads to the cumulative energy contributions from ${\Phi}_{a}$ ($\approx$ 2.7) being nearly constant across the inertial region (figure \ref{fig5}(a)).
Such a trend is consistent with the statistical properties of the active motions scaling universally with $U_{\tau}$ and $z$ \citep{townsend1961,bradshaw1967}.

\subsection{Inactive contributions from the self-similar attached eddies}
\label{inactive_AE}

\begin{figure}
\begin{center}
\includegraphics[width=1.0\textwidth]{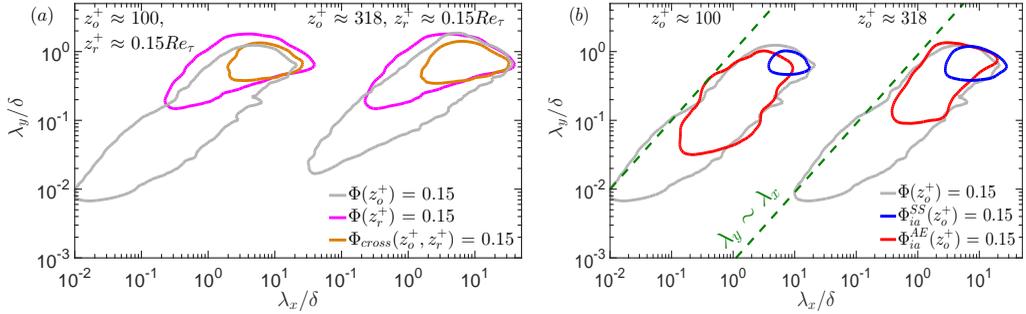}
\caption{(a) Constant energy contours for ${\Phi}$($z^{+}_{o}$), ${\Phi}_{cross}$($z^{+}_{o}$, $z^{+}_{r}$ $\approx$ 0.15${Re_{\tau}}$) and ${\Phi}$($z^{+}_{r}$ $\approx$ 0.15$Re_{\tau}$) at energy level of 0.15 plotted for $z^{+}_{o}$ $\approx$ 100 and 318. 
(b) Constant energy contours for ${\Phi}^{AE}_{ia}$($z^{+}_{o}$) and ${\Phi}^{SS}_{ia}$($z^{+}_{o}$), computed via (\ref{eq9}) and (\ref{eq10}), plotted at the same energy level and $z^{+}_{o}$ as in (a). 
All contours in (a,b) are computed from the high $Re_{\tau}$ experimental data. 
Dashed green lines represent the linear relationship, ${\lambda}_{y}$ $\sim$ ${\lambda}_{x}$.}
\label{fig6}
\end{center}
\end{figure}

Here, we consider isolating the inactive contributions from the self-similar attached eddies, by first estimating the $\delta$-scaled superstructure contributions to ${\Phi}$($z_{o}$).
As discussed previously in $\S$\ref{intro} and observed from the experimental data in figures \ref{fig2} and \ref{fig4}, the superstructures extend from the wall and span across the entire inertial region, while contribution from the tallest attached eddies is insignificant beyond the upper bound of the log-region ($z^+$ $\sim$ 0.15$Re_{\tau}$).
This is supported by the scale-by-scale coupling (${\Phi}_{cross}$) computed from the $u$-signals simultaneously measured at $z^{+}_{o}$ ($\approx$ 100, 200 or 318) and $z^{+}_{r}$ $\approx$ 0.15$Re_{\tau}$ plotted in figure \ref{fig6}(a), where the energy contours can be seen to be restricted only to the very large scale end of ${\Phi}$, indicative of the superstructure signature.
The choice of $z^{+}_{r}$ $\approx$ 0.15$Re_{\tau}$ is also consistent with \citet{baars2020,baars2020part2}, who also used it as a reference location to extract the superstructure contribution.  
They recommended keeping $z^{+}_{o}$ $\lesssim$ ${z^{+}_{r}}/8$ to meet the requirement of $z^{+}_{r}$ $\gg$ $z^{+}_{o}$, which explains the present ${\Phi}_{cross}$($z^{+}_{o},z^{+}_{r}$ $\approx$ 0.15$Re_{\tau}$) measurements conducted at only three wall-normal locations ($z^{+}_{o}$) in the inertially dominated region.
It is worth noting here that owing to this condition, the cross-spectrum analysis to isolate the superstructure contribution is only possible on the high $Re_{\tau}$ experimental dataset.
On computing ${\Phi}_{cross}$($z^{+}_{o}$,$z^{+}_{r}$ $\approx$ 0.15$Re_{\tau}$) from the experimental data, it is used in conjunction with the SLSE (appendix 1) to obtain a linear stochastic estimate of the spectrum (${\Phi}^{SS}_{ia}$) associated with the superstructure contributions at $z_{o}$ following:
\begin{equation}
\label{eq9}
{{{\Phi}^{SS}_{ia}}(z^{+}_{o};{\lambda_{x}},{\lambda_{y}})} = { \frac{ {\left[{{\Phi}_{cross}}({z^{+}_{o}}, {z^{+}_{r}} \approx 0.15{Re_{\tau}};{\lambda_{x}},{\lambda_{y}}) \right]}^{2} }{ {{\Phi}}({z^{+}_{r}} \approx 0.15{Re_{\tau}};{\lambda_{x}},{\lambda_{y}}) } }.
\end{equation}
The above expression is similar to (\ref{eq7}) discussed in $\S$\ref{slse}, with the calculations in (\ref{eq9}) also carried out first in the frequency domain, followed by the conversion to ${\lambda}_{x}$ done by invoking Taylor's hypothesis using $U_{c}$ $=$ $U$($z^{+}_{r}$ $\approx$ 0.15$Re_{\tau}$) \citep{baars2020,baars2020part2}.
The choice of $U_{c}$ in (\ref{eq9}) is based on the `global' nature and high convection speeds of the $\delta$-scaled superstructures \citep{jimenez2008,del2009,monty2009dns}.

\begin{figure}
\begin{center}
\includegraphics[width=0.9\textwidth]{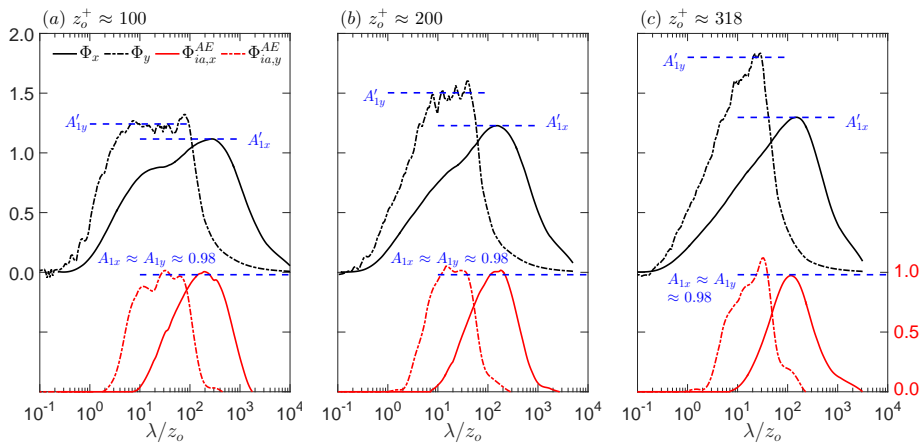}
\caption{${\Phi}$($z^{+}_{o}$) and ${\Phi}^{AE}_{ia}$($z^{+}_{o}$) integrated across ${\lambda}_{y}$ and ${\lambda}_{x}$ to obtain their corresponding premultiplied 1-D version as a function of ${\lambda}_{x}$ (${\Phi}_{x}$, ${\Phi}^{AE}_{ia,x}$; in solid line) and ${\lambda}_{y}$ (${\Phi}_{y}$, ${\Phi}^{AE}_{ia,y}$ ; in dash-dotted line), respectively, for $z^{+}_{o}$ $\approx$ (a) 100, (a) 200 and (c) 318. 
Also highlighted are the peaks/plateaus of ${\Phi}^{AE}_{ia,x}$ and ${\Phi}^{AE}_{ia,y}$ ($A_{1x}$, $A_{1y}$), along with those of ${\Phi}_{x}$ and ${\Phi}_{y}$ ($A'_{1x}$, $A'_{1y}$).}
\label{fig7}
\end{center}
\end{figure}

Contours associated with all the energy spectra in (\ref{eq9}) have been plotted in figure \ref{fig6}(b), with ${\Phi}^{SS}_{ia}$ centred around a $\delta$-scaled location of ${\lambda}_{x}$ $\sim$ 7$\delta$, ${\lambda}_{y}$ $\sim$ 0.7$\delta$, representative of the superstructures. 
Following the linear superposition assumption in (\ref{eq3a}), ${\Phi}^{SS}_{ia}$($z_{o}$) can be simply subtracted from ${\Phi}_{ia}$($z_{o}$) to estimate the inactive contributions from the attached eddies at $z_{o}$ (${\Phi}^{AE}_{ia}$):
\begin{equation}
\label{eq10}
{{{\Phi}^{AE}_{ia}}(z^{+}_{o};{\lambda_{x}},{\lambda_{y}})} = {{{\Phi}_{ia}}(z^{+}_{o};{\lambda_{x}},{\lambda_{y}})} - {{{\Phi}^{SS}_{ia}}(z^{+}_{o};{\lambda_{x}},{\lambda_{y}})},
\end{equation}
with ${\Phi}^{SS}_{ia}$ and ${\Phi}^{AE}_{ia}$ representative of $\overline{u^{2}}^{+}_{\rm inactive,SS}$ and $\overline{u^{2}}^{+}_{\rm inactive,AE}$, respectively.
The claim is also supported by the ${\Phi}^{AE}_{ia}$ contours plotted in figure \ref{fig6}(b), which are seen to follow the ${\lambda}_{y}$ $\sim$ ${\lambda}_{x}$ relationship representing geometric self-similarity.
$\overline{u^{2}}^{+}_{\rm inactive,AE}$, obtained via integrating ${\Phi}^{AE}_{ia}$ at the three $z^{+}_{o}$, has also been plotted in figure \ref{fig5}(b).
While the trend looks promising when compared with the ${\overline{u^2}}^{+}$ expression in (\ref{eq1}), three data points are not sufficient to firmly establish the present claim (for this additional data at even higher Reynolds number would be required).
However, to confirm the association of ${\Phi}^{AE}_{ia}$ with pure attached eddy contributions, we check for the constant energy plateau (representative of the $k^{-1}_{x}$-scaling) in the corresponding premultiplied 1-D spectra.
In this respect, the parameters in the present analysis align well with the necessary conditions proposed by \citet{perry1982} and \citet{nickels2005} to observe a clear $k^{-1}_{x}$ region, i.e. to measure sufficiently close to the wall in a high $Re_{\tau}$ wall-bounded flow.

To this end, ${\Phi}^{AE}_{ia}$ is integrated along ${\lambda}_{y}$ and ${\lambda}_{x}$ to obtain the corresponding premultiplied 1-D spectra as a function of ${\lambda}_{x}$ (${\Phi}^{AE}_{ia,x}$) and ${\lambda}_{y}$ (${\Phi}^{AE}_{ia,y}$), respectively, which has been plotted for the three $z^{+}_{o}$ in figure \ref{fig7}.
Also plotted for reference in the same figure are the premultiplied 1-D spectra (${\Phi}_{x}$,${\Phi}_{y}$) obtained by integrating ${\Phi}$ in the same manner.
Indeed, ${\Phi}^{AE}_{ia,x}$ can be observed to be plateauing at $A_{1x}$ $\approx$ 0.98 (nominally) for all three $z^{+}_{o}$, which is consistent with the Townsend-Perry constant ($A_{1}$) estimated by \citet{baars2020part2} from the streamwise turbulence intensity profile.
The span of the ${\Phi}^{AE}_{ia,x}$ plateau, however, shrinks in size with the increase in $z^{+}_{o}$, likely due to decrease in the hierarchy of attached eddies inactive at $z_{o}$ \citep{perry1982}.
To the best of the authors' knowledge, the present result is the first empirical evidence that establishes consistency between the logarithmic decay rate of the streamwise turbulence intensity and the constant energy plateau from the premultiplied 1-D $u$-spectrum, as argued by \citet{perry1986} in the case of pure attached eddy contributions.
This consistency, however, was not observed in the recent effort by \citet{baars2020,baars2020part2} due to the energy decomposition conducted directly for the 1-D $u$-spectra in their case.
That analysis neglected the scale-specific coherence over the spanwise direction, which has been duly considered in the present study using the new experimental data.

The present analysis also reveals the plateau, $A_{1y}$ in the premultiplied spanwise 1-D spectra (${\Phi}^{AE}_{ia,y}$), which is found to be nominally equal to $A_{1x}$ for all $z^{+}_{o}$.
$A_{1y}$ $\approx$ $A_{1x}$ obtained here, thus satisfies the necessary condition proposed by \citet{chandran2017} to associate the 2-D spectrum, ${\Phi}^{AE}_{ia}$ with purely self-similar contributions.
${\Phi}_{y}$ is also observed to have a plateau at $A'_{1y}$ $\approx$ 1.3 (at $z^{+}_{o}$ $\approx$ 100), a value which is consistent with that reported by \citet{mklee2015}.
However, $A'_{1y}$ $\neq$ $A'_{1x}$, with both values changing as a function of $z^{+}_{o}$.
This behaviour can be associated with the non-self-similar contributions in $\Phi$ obscuring the pure self-similar characteristics, which have been successfully isolated in the present study in the form of ${\Phi}^{AE}_{ia}$.

\section{Concluding remarks}

The present study proposes a methodology to extract the $u$-energy spectrum associated with the active and inactive motions \citep{townsend1961,townsend1976} coexisting at any $z_{o}$ in the inertially-dominated region of a wall-bounded flow.
The methodology is based on isolating the streamwise turbulent energy associated with the inactive motions from the total energy, 
based on their known characteristic of being larger than the coexisting active motions and coherent across a substantial wall-normal range \citep{townsend1961,townsend1976}.
This is tested using ZPG TBL datasets comprising two-point $u$-signals, synchronously acquired at $z_{o}$ and a near-wall location ($z_{r}$), such that $z_{r}$ $\ll$ $z_{o}$.
The velocity-velocity coupling, constructed by cross-correlating these $u$-signals, is fed into an SLSE-based procedure which linearly decomposes the full 2-D spectrum $\Phi$($z_{o}$) into components representative of the active (${\Phi}_{a}$) and inactive (${\Phi}_{ia}$) motions at $z_{o}$.

${\Phi}_{a}$ is found to exhibit $z_{o}$-scaling across a decade of $Re_{\tau}$, and is also consistent with the characteristics depicted by the Reynolds shear stress cospectra, thereby confirming the association of ${\Phi}_{a}$ with the active motions.
Analysis conducted across both spatially-(DNS) and temporally-resolved (experimental) datasets also confirms the validity of Taylor's hypothesis for the active motions.
Further, decomposition of ${\Phi}$ into ${\Phi}_{a}$ and ${\Phi}_{ia}$ brings out the self-similar characteristic of the two spectra, which is consistent with Townsend's hypothesis on both active and inactive motions essentially being associated with contributions from the attached eddies, but of different sizes.
In terms of usefulness to reduced-order modelling, the present study highlights the close match between the geometry of the active motions and the self-similar vortex clusters investigated previously \citep{del2006,hwang2015}, suggesting the former could be modelled along similar lines as the latter.

While ${\Phi}_{a}$($z_{o}$) is found to be associated predominantly with the self-similar attached eddies of height $\mathcal{H}$ $\sim$ $\mathcal{O}(z_{o})$, ${\Phi}_{ia}$($z_{o}$) is found to have contributions from both, the relatively tall self-similar attached eddies ($\mathcal{O}(z_{o})$ $\ll$ $\mathcal{H}$ $\lesssim$ $\mathcal{O}(\delta)$) as well as the large $\delta$-scaled eddies associated with the superstructures.
The latter is confirmed by the reduced self-similar contributions to ${\Phi}_{ia}$ with increasing $z_{o}$, due to the large attached eddies qualifying as active in accordance to the original concept given by \citet{townsend1961,townsend1976}.
The present study also segregates the inactive contributions from the attached eddies (${\Phi}^{AE}_{ia}$), from those coming from the $\delta$-scaled superstructures (${\Phi}^{SS}_{ia}$), by utilizing the same SLSE-based methodology used earlier.
The estimation of ${\Phi}^{AE}_{ia}$ reveals the constant energy plateau, representative of $k^{-1}$-scaling, in the corresponding premultiplied streamwise and spanwise 1-D $u$-spectra.
Both these spectra were found to plateau at $A_{1}$ $\approx$ 0.98 (nominally), yielding the first empirical evidence to establish the consistency with $A_{1}$ obtained from the streamwise turbulence intensity profiles \citep{baars2020part2}, as argued by \citet{perry1982} for the case of pure attached eddy contributions.

\section*{Acknowledgements}

The authors wish to acknowledge the Australian Research Council for financial support and are thankful to the authors of \citet{sillero2014} and \citet{baidya2017} for making their respective data available. The authors also thank Dr. D. Chandran for assistance with the experiments, and Dr. W. J. Baars and Dr. A. Madhusudanan for helpful discussions related to SLSE.
The authors are also grateful to the anonymous reviewers for their helpful comments which significantly improved the quality of the manuscript.

\section*{Declaration of Interests} 

The authors report no conflict of interest.

\section*{Appendix 1: SLSE methodology adopted for energy decomposition}
\label{app1}

Here, we demonstrate the methodology to estimate a component of the full $u$-energy spectrum at $z_{o}$, comprising contributions from specific coherent motions coexisting at $z_{o}$, via the spectral linear stochastic estimation (SLSE) approach.
The procedure has been adopted from previous studies in the literature employing SLSE \citep{tinney2006,baars2016slse,anagha2019,encinar2019}, which may be referred to for further understanding on this topic.
The SLSE considers a scale-specific unconditional input (at $z_{r}$) to give a scale-specific conditional output (at $z_{o}$) following:
\begin{equation}
\label{app_eq9}
{\widetilde{u}}^{E}(z_{o};{\lambda_{x}},{\lambda_{y}}) = {{H_{L}}(z_{o},z_{r};{\lambda_{x}},{\lambda_{y}})}\widetilde{u}(z_{r};{\lambda_{x}},{\lambda_{y}}), 
\end{equation}
where ${\tilde{u}}(z_{r};{\lambda_{x}},{\lambda_{y}})$ is the 2-D Fourier transform of $u({z_{r}})$ in $x$ and $y$.
Here, the superscript $E$ represents the estimated quantity and $H_{L}$ represents the scale-specific linear transfer kernel.
It should be noted that the SLSE approach enables accurate estimation of only those scales (at $z_{o}$) that are coherent across $z_{o}$ and $z_{r}$. 
Equation (\ref{app_eq9}) can be further used to estimate the 2-D energy spectrum, ${\Phi}^{E}$ at $z_{o}$ \citep{anagha2019} following:
\begin{equation}
\label{app_eq10}
{{\Phi}^{E}}(z_{o};{\lambda_{x}},{\lambda_{y}}) = {{\mid {H_{L}}(z_{o},{z_{r}};{\lambda_{x}},{\lambda_{y}}) \mid}^{2}}{{\Phi}({z_{r}};{\lambda_{x}},{\lambda_{y}})}. 
\end{equation}
To obtain $u^{E}$ and ${\Phi}^{E}$ at $z_{o}$, the transfer kernel $H_{L}$ is required to be computed from an ensemble of data following:
\begin{equation}
\label{app_eq11}
{H_{L}}(z_{o},z_{r};{\lambda_{x}},{\lambda_{y}}) = \frac{ {\langle \widetilde{u}(z_{o};{\lambda_{x}},{\lambda_{y}}){{\widetilde{u}}^{\ast}}(z_{r};{\lambda_{x}},{\lambda_{y}}) \rangle} }{ {\langle {\widetilde{u}(z_{r};{\lambda_{x}},{\lambda_{y}})}{{\widetilde{u}}^{\ast}(z_{r};{\lambda_{x}},{\lambda_{y}})} \rangle} } = {\mid {{H_{L}}(z_{o},z_{r};{\lambda_{x}},{\lambda_{y}})} \mid}{e^{i{\psi}(z_{o},z_{r};{\lambda_{x}},{\lambda_{y}})}},
\end{equation}
with ${\mid {H_{L}} \mid}$ and $\psi$ the scale-specific gain and phase respectively, and the asterisk ($\ast$), angle brackets ($\langle \rangle$) and vertical bars ($\mid \mid$) denoting the complex conjugate, ensemble averaging and modulus, respectively.
Considering $z^{+}_{r}$ as the reference wall-normal location used in the present study, ${\mid {H_{L}} \mid}$ from equation (\ref{app_eq11}) can be simply expressed as a function of the two types of 2-D spectra computed from the multi-point datasets (refer $\S$\ref{data}) at various $z^{+}_{o}$ in the inertial region following:
\begin{equation}
\label{app_eq12} 
{\mid {{H_{L}}(z^{+}_{o},{z^{+}_{r}};{\lambda_{x}},{\lambda_{y}})} \mid} = \frac{{{\Phi}_{cross}}({z^{+}_{o}},{z^{+}_{r}};{\lambda_{x}},{\lambda_{y}})}{{\Phi}({z^{+}_{r}};{\lambda_{x}},{\lambda_{y}})}.
\end{equation}

In the case of $z^{+}_{r}$ $\ll$ $z^{+}_{o}$, ${{{\Phi}^{E}}(z_{o};{\lambda_{x}},{\lambda_{y}})}$ would be representative of the energy contributions from all coexisting motions taller than $z_{o}$, which as per our discussion in $\S$\ref{slse} leads to ${{{{\Phi}^{E}}(z_{o};{\lambda_{x}},{\lambda_{y}})}{{\big|}_{{z^{+}_{r}} \approx 15}}}$ ${\rightarrow}$ ${{\Phi}_{ia}}(z_{o};{\lambda_{x}},{\lambda_{y}})$.
A simplified expression for ${\Phi}_{ia}$ can be deduced from (\ref{app_eq10}) and (\ref{app_eq12}) as follows: 
\begin{equation}
\label{app_eq13}
\begin{aligned}
{{\Phi}_{ia}}(z^{+}_{o};{\lambda_{x}},{\lambda_{y}}) &= {{\mid {H_{L}}(z^{+}_{o},{z^{+}_{r}} \approx 15;{\lambda_{x}},{\lambda_{y}}) \mid}^{2}}{{{\Phi}}({z^{+}_{r}} \approx 15;{\lambda_{x}},{\lambda_{y}})} \\ &= { \frac{ {\left[{{\Phi}_{cross}}({z^{+}_{o}},{z^{+}_{r}} {\approx} 15;{\lambda_{x}},{\lambda_{y}}) \right]}^{2} }{ {{\Phi}}({z^{+}_{r}} \approx 15;{\lambda_{x}},{\lambda_{y}}) } }. 
\end{aligned}
\end{equation}
Similarly, in case of $z^{+}_{r}$ $\gg$ $z^{+}_{o}$, ${{{\Phi}^{E}}(z_{o};{\lambda_{x}},{\lambda_{y}})}$ would be representative of the energy contributions from all coexisting motions at $z_{o}$ that are taller than $z_{r}$, which as per our discussion in $\S$\ref{inactive_AE} leads to ${{{{\Phi}^{E}}(z_{o};{\lambda_{x}},{\lambda_{y}})}{{\big|}_{{z^{+}_{r}} \approx 0.15Re_{\tau}}}}$ ${\rightarrow}$ ${{\Phi}^{SS}_{ia}}(z_{o};{\lambda_{x}},{\lambda_{y}})$, and can be estimated following: 
\begin{equation}
\label{app_eq13a}
\begin{aligned}
{{\Phi}^{SS}_{ia}}(z^{+}_{o};{\lambda_{x}},{\lambda_{y}}) = { \frac{ {\left[{{\Phi}_{cross}}({z^{+}_{o}},{z^{+}_{r}} \approx 0.15{Re_{\tau}};{\lambda_{x}},{\lambda_{y}}) \right]}^{2} }{ {{\Phi}}({z^{+}_{r}} \approx 0.15Re_{\tau};{\lambda_{x}},{\lambda_{y}}) } }. 
\end{aligned}
\end{equation}
Availability of both the numerator and denominator in the above expressions (table \ref{tab_exp}) allows direct computation of ${\Phi}_{ia}$ and ${\Phi}^{SS}_{ia}$, without separately estimating ${\mid H_{L} \mid}^2$, for both the datasets.
It should be noted here that ${\Phi}_{ia}$ is computed in the frequency domain for the experimental dataset, with the conversion to ${\lambda}_{x}$ obtained by invoking Taylor's hypothesis, using $U_{c}$ $=$ $U$($z_{o}$) \citep{baars2016slse,baars2017}.
${\Phi}^{SS}_{ia}$ is also computed in the similar manner, however with the conversion to ${\lambda}_{x}$ achieved by using $U_{c}$ $=$ $U$($z^{+}_{r}$ $\approx$ 0.15$Re_{\tau}$) \citep{baars2020,baars2020part2}.

\begin{figure}
\begin{center}
\includegraphics[width=0.9\textwidth]{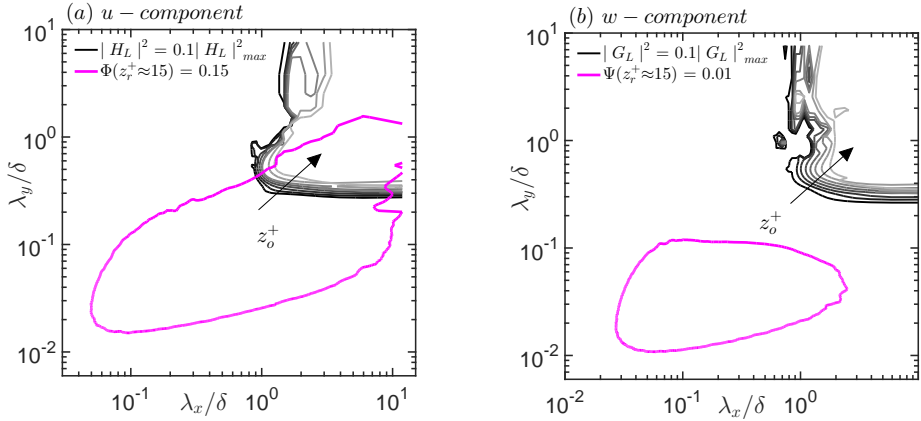}
\caption{Contours for the linear transfer kernel (black), for $z^{+}_{r}$ $\approx$ 15, and the near-wall 2-D spectra (magenta) for the (a) $u$- and (b) $w$-velocity component. 
Here, the transfer kernels are computed at various $z^{+}_{o}$ from the DNS dataset of \citet{sillero2014}, described in table \ref{tab_exp}. 
Dark to light shading indicates an increase in $z^{+}_{o}$ following 100 $\lesssim$ $z^{+}_{o}$ $\lesssim$ 0.15$Re_{\tau}$.
The contour levels for the transfer kernels, ${\mid H_{L} \mid}^2$ and ${\mid G_{L} \mid}^2$ correspond to approximately 10\% of the maximum value recorded for the kernel at the respective $z^{+}_{o}$, while that for ${\Psi}$($z^{+}_{r}$ $\approx$ 15) has intentionally been kept very low to highlight no overlap with the associated transfer kernel, ${\mid G_{L} \mid}^2$.}
\label{fig1_app}
\end{center}
\end{figure} 

As can be noted from (\ref{app_eq13}), the essential information on energetic motions coherent across $z_{o}$ and $z_{r}$ is embedded in ${\mid H_{L} \mid}^2$ which is translated into ${\Phi}_{ia}$ (or ${\Phi}^{SS}_{ia}$) via scale-by-scale amplification/attenuation provided by ${\Phi}$($z^{+}_{r}$).
For example, figure \ref{fig1_app}(a) shows the ${\mid {H_{L}}(z^{+}_{o},{z^{+}_{r}} {\approx} 15) \mid}^2$ computed from the DNS dataset at various $z^{+}_{o}$ listed in table \ref{tab_exp}, along with ${\Phi}$($z^{+}_{r}$ $\approx$ 15) for the same dataset.
It is evident from the plot that ${\mid{H_{L}}(z^{+}_{o} {\approx} 100,{z^{+}_{r}} {\approx} 15)\mid}^2$ contours conform predominantly to the large scales of the spectrum, with the contours moving very gradually to even larger scales with increase in $z^{+}_{o}$.
This explains the observation noted in $\S$\ref{slse} on ${\Phi}_{ia}$ and ${\Phi}_{a}$ respectively taking up the higher and lower end of the full spectrum, ${\Phi}$($z_{o}$).

We also use this opportunity to test the efficacy of the SLSE-based methodology, used here to estimate energy contributions from the active motions.
To this end, the same procedure as that outlined in (\ref{app_eq9}) -- (\ref{app_eq13}), is implemented on the wall-parallel $w$-velocity fields also retrieved from the DNS dataset, at the same $z^{+}_{o}$ and $z^{+}_{r}$ as that selected for the $u$-velocity field (table \ref{tab_exp}).
The $w$-velocity fields are used to compute the two types of 2-D spectra, computed previously for the $u$-component, at the same $z^{+}_{o}$ in the inertial region following:
\begin{equation}
\label{app_eq14}
\begin{split}
{\Psi}(z^{+}_{o};{{\lambda}_{x}},{{\lambda}_{y}}) &= {\mid}{k^{+}_{x}}{k^{+}_{y}}{{\phi}^{+}_{{w_{o}}{w_{o}}}}({z^{+}_{o}};{{\lambda}_{x}},{{\lambda}_{y}}){\mid} \; \text{and}\\
{{\Psi}_{cross}}(z^{+}_{o},{z^{+}_{r}} {\approx} 15;{{\lambda}_{x}},{{\lambda}_{y}}) &= {\mid}{k^{+}_{x}}{k^{+}_{y}}{{\phi}^{+}_{{w_{o}}{w_{r}}}}({z^{+}_{o}},{z^{+}_{r}}\;{\approx}\;15;{{\lambda}_{x}},{{\lambda}_{y}}){\mid},
\end{split}
\end{equation}  
where (\ref{app_eq14}) is analogous to (2.3) expressed for the $u$-component.
Similarly, the inactive component ${\Psi}_{ia}$ of the $w$-velocity field can be computed via an expression similar to (\ref{app_eq13}) given previously for the $u$-component:
\begin{equation}
\label{app_eq15}
\begin{aligned}
{{\Psi}_{ia}}(z^{+}_{o};{\lambda_{x}},{\lambda_{y}}) &= {{\mid {G_{L}}(z^{+}_{o},{z^{+}_{r}} \approx 15;{\lambda_{x}},{\lambda_{y}}) \mid}^{2}}{{{\Psi}}({z^{+}_{r}} \approx 15;{\lambda_{x}},{\lambda_{y}})}, \; \text{where}\\ 
{\mid {{G_{L}}(z^{+}_{o},{z^{+}_{r}} \approx 15;{\lambda_{x}},{\lambda_{y}})} \mid} &= \frac{{{\Psi}_{cross}}({z^{+}_{o}},{z^{+}_{r}} {\approx} 15;{\lambda_{x}},{\lambda_{y}})}{{\Psi}({z^{+}_{r}} \approx 15;{\lambda_{x}},{\lambda_{y}})}. 
\end{aligned}
\end{equation}
Accordingly, ${\Psi}_{a}$($z_{o}$) $=$ ${\Psi}$($z_{o}$) -- ${\Psi}_{ia}$($z_{o}$). 
Given the fact that the $w$-component is predominantly associated with the active motions \citep{bradshaw1967,morrison1992,baidya2017}, i.e. ${\Psi}_{a}$($z_{o}$) $\approx$ $\Psi$($z_{o}$), we would expect the SLSE procedure to reveal cumulative energy contributions from ${\Psi}_{ia}$ to be negligible.
This is possible if ${{\mid {G_{L}} \mid}^2}$ and ${\Psi}$($z^{+}_{r}$ $\approx$ 15) do not overlap at common scales (as per equation \ref{app_eq15}).
Figure \ref{fig1_app}(b) shows ${{\mid {G_{L}} \mid}^2}$ contours computed for various $z^{+}_{o}$ from the DNS dataset alongside ${\Psi}$($z^{+}_{r}$ $\approx$ 15), clearly suggesting an insignificant overlap between the two.
Accordingly, ${\Psi}_{ia}$($z_{o}$) $\approx$ 0 in the inertial region leading to ${\Psi}_{a}$($z_{o}$) $\approx$ ${\Psi}$($z_{o}$), which proves the effectiveness of the SLSE-based methodology in extracting the energy spectrum associated with the active motions.

\bibliographystyle{jfm}
\bibliography{Active_inactive_uMotions_bib}

\end{document}